\documentclass[conference,letterpaper]{IEEEtran}

\usepackage[utf8]{inputenc} 
\usepackage[T1]{fontenc}
\usepackage{url}
\usepackage{ifthen}
\usepackage{cite}
\usepackage{graphicx}
\usepackage{xcolor}
\usepackage{hyperref}
\usepackage[cmex10]{amsmath} % Use the [cmex10] option to ensure complicance
\DeclareMathOperator*{\argminB}{argmin}   % Jan Hlavacek

\usepackage{makecell}
\setcellgapes{2pt}
\usepackage[thinlines]{easytable}
\newcommand{\floor}[1]{\left\lfloor #1 \right\rfloor}

\def\BibTeX{{\rm B\kern-.05em{\sc i\kern-.025em b}\kern-.08em
    T\kern-.1667em\lower.7ex\hbox{E}\kern-.125emX}}
\ifCLASSOPTIONcompsoc
    \usepackage[caption=false, font=normalsize, labelfont=sf, textfont=sf]{subfig}
\else
\usepackage[caption=false, font=footnotesize]{subfig}
\fi
\usepackage{lipsum}
\begin{document}
\title{Neural-Network-Optimized Degree-Specific Weights for LDPC MinSum Decoding} 

% %%% Single author, or several authors with same affiliation:
% \author{%
%   \IEEEauthorblockN{Stefan M.~Moser}
%   \IEEEauthorblockA{ETH Zürich\\
%                     ISI (D-ITET)\\
%                     CH-8092 Zürich, Switzerland\\
%                     Email: moser@isi.ee.ethz.ch}
% }

%%% Several authors with up to three affiliations:
\author{%
  \IEEEauthorblockN{Linfang Wang*, Sean Chen*, Jonathan Nguyen*, Divsalar Dariush\dag , Richard Wesel*}
  \IEEEauthorblockA{*Department of Electrical and Computer Engineering, University of California, Los Angeles, Los Angeles, California 90095\\
                    \dag Jet Propulsion Laboratory, California Institute of Technology, Pasadena, California 91109\\
                    Email: \{lfwang, nguyen.j,wesel\}@ucla.edu, mistystory@g.ucla.edu, Dariush.Divsalar@jpl.nasa.gov}
  %\and
  %\IEEEauthorblockN{Albus Dumbledore and Harry Potter}
  %\IEEEauthorblockA{Hogwarts School of Witchcraft and Wizardry\\
                    %Hogwarts Castle\\ 
                    %1714 Hogsmeade, Scotland\\
                    %Email: \{dumbledore, potter\}@hogwarts.edu}
}

%%% Many authors with many affiliations:
% \author{%
%   \IEEEauthorblockN{Albus Dumbledore\IEEEauthorrefmark{1},
%                     Olympe Maxime\IEEEauthorrefmark{2},
%                     Stefan M.~Moser\IEEEauthorrefmark{3}\IEEEauthorrefmark{4},
%                     and Harry Potter\IEEEauthorrefmark{1}}
%   \IEEEauthorblockA{\IEEEauthorrefmark{1}%
%                     Hogwarts School of Witchcraft and Wizardry,
%                     1714 Hogsmeade, Scotland,
%                     \{dumbledore, potter\}@hogwarts.edu}
%   \IEEEauthorblockA{\IEEEauthorrefmark{2}%
%                     Beauxbatons Academy of Magic,
%                     1290 Pyrénées, France,
%                     maxime@beauxbatons.edu}
%   \IEEEauthorblockA{\IEEEauthorrefmark{3}%
%                     ETH Zürich, ISI (D-ITET), ETH Zentrum, 
%                     CH-8092 Zürich, Switzerland,
%                     moser@isi.ee.ethz.ch}
%   \IEEEauthorblockA{\IEEEauthorrefmark{4}%
%                     National Chiao Tung University (NCTU), 
%                     Hsinchu, Taiwan,
%                     moser@isi.ee.ethz.ch}
% }

\maketitle

%%%%%%
%% Abstract: 
%% If your paper is eligible for the student paper award, please add
%% the comment "THIS PAPER IS ELIGIBLE FOR THE STUDENT PAPER
%% AWARD." as a first line in the abstract. 
%% For the final version of the accepted paper, please do not forget
%% to remove this comment!
%%
\begin{abstract}
Neural Normalized MinSum (N-NMS) decoding delivers better frame error rate (FER)  performance on linear block codes than conventional normalized MinSum (NMS) by assigning dynamic multiplicative weights to each check-to-variable message in each iteration. Previous N-NMS efforts have  primarily investigated short-length block codes ($N<1000$), because the number of N-NMS parameters to be trained is proportional to the number of edges in the parity check matrix and the number of iterations, which imposes am impractical memory requirement when Pytorch or Tensorflow is used for training.
This paper provides efficient approaches to training parameters of N-NMS that support N-NMS for longer block lengths. 
Specifically, this paper introduces a family of  neural 2-dimensional normalized (N-2D-NMS) decoders with  with various reduced parameter sets and shows how performance varies with the parameter set selected. 
The N-2D-NMS decoders share weights with respect to check node and/or variable node degree. 
Simulation results justify this approach, showing that the trained weights of N-NMS have a strong correlation to the check node degree, variable node degree, and iteration number. 
Further simulation results on the (3096,1032) protograph-based raptor-like (PBRL) code show that N-2D-NMS decoder can achieve the same FER as N-NMS with significantly fewer parameters required. 
The N-2D-NMS decoder for a (16200,7200) DVBS-2 standard LDPC code shows a lower error floor than belief propagation.
Finally, a hybrid decoding structure combining a feedforward structure with a recurrent structure is proposed in this paper. The hybrid structure shows similar decoding performance to full feedforward structure, but requires significantly fewer parameters.

\end{abstract}

\section{Introduction}

{\let\thefootnote\relax\footnote{{This research is supported by Physical Optics Corporation (POC), SA Photonics, and National Science Foundation (NSF) grant CCF-1911166. Any opinions, findings, and conclusions or recommendations expressed in this material are those of the author(s) and do not necessarily reflect views of POC, SA or NSF. Research was carried out in part at the Jet Propulsion Laboratory, California Institute of Technology, under a contract with NASA. \copyright 2021. All rights reserved.}}}

Message passing decoders are often used for linear block code decoding. Typical message passing decoders utilize belief propagation (BP), MinSum, and its variations such as normalized MinSum (NMS) and offset MinSum (OMS). However, message passing decoders are sub-optimal because of the existence of cycles in the parity check matrix. 

Recently, numerous works have focused on enhancing the performance of message passing decoders with the help of neural networks \cite{Nachmani2016-bs,Lugosch2017-ed,Nachmani2017-qq,Nachmani2018-ra,Liang2018-lw,Wu2018-zr,Lugosch2018-gu,Lyu2018-nz,Xiao2019-kj,Deng2019-cf,Abotabl2019-wt,Buchberger2020-pf,Wang2020-fb,Lian2019-jh}. The neural network is created by unfolding the message passing operations of each decoding iteration \cite{Nachmani2016-bs}.
%Each decoding iteration is unfolded into two hidden layers which represent check node processing layer and variable node processing layer and each neuron represents a variable-to-check message or a check-to-variable message. 
Nachmani \textit{et al.} in \cite{Nachmani2016-bs} proposed improving BP decoding by assigning unique multiplicative weights to check-to-variable messages and the channel log-likelihood (LLR) of variables in each iteration. The so-called "Neural BP (NBP)" has shown better performance than BP. The authors further proposed a recurrent neural network  BP (RNN-BP) \cite{Nachmani2017-qq} decoder, which set the edge-specific weight to be equal in each iteration. Nachmani \textit{et al.} and  Lugosch \textit{et al.} in \cite{Nachmani2018-ra, Lugosch2017-ed,Nachmani2016-bs} proposed a Neural Normalized MinSum (N-NMS) decoder and Neural Offset MinSum (N-OMS) decoder to improve the performance of the NMS and OMS decoder. 

As the code length gets longer, these edge-specific neural decoders become impractical because the number of edges scales quickly. One solution is to share one parameter with edges that have same properties, such as in the same iteration, or connecting same check/variable node. For an example, Wang \textit{et al.} proposed to assign parameters for each check-to-variable layer and variable-to-check layer \cite{Wang2020-fb}, respectively. M. Lian  \emph{et. al.} in \cite{Lian2019-jh}  considered assigning same weight to all messages in one iteration. 

Besides, most previous work has focused on codes with short block lengths ($N<1000$). The focus on short codes may result from the fact that popular deep learning research platforms, such as Pytorch and Tensorflow, require large amounts of memory to calculate the gradient when the block length is long. However, as pointed in \cite{Abotabl2019-wt}, it is possible to train the parameters for longer block lengths if resources are handled more efficiently. Abotabl \textit{et al.} provided an efficient computation framework for optimizing the offset values in the N-OMS algorithm\cite{Abotabl2019-wt}, and trained an OMS neural network with edge-specific weights, iteration-specific weights, and a single weight. 

A primary contribution of this paper is a family of neural 2-dimensional normalized MinSum (N-2D-NMS) decoder whose weights are optimized by a neural network based on node degree.  This simplification over previous approaches that separately optimize the weight for each edge leads to a much simpler optimization that provides excellent FER performance while still accommodating large block lengths of practical importance. The main contributions in this paper are:

\begin{itemize}
    \item An efficient implementation of the N-NMS architecture based on the gradients in backpropagation. This part is related to the framework in \cite{Abotabl2019-wt}. We show that backpropagation can be conducted in an iterative way, and the parameters used to implement backpropagation can be stored efficiently. This approach solves the memory issue faced by Tensorflow \cite{Lugosch_undated-wi} and facilitates  an efficient C++ implementation simple enough to train without a GPU.
    \item Empirical N-NMS results  demonstrating that dynamic weights show a strong correlation with check node degree, variable node degree, and iteration.
    \item A family of N-2D-NMS decoders with various reduced parameter sets showing how performance varies with the parameter set selected.   The N-2D-NMS decoding structure  is a generalization of  \cite{Juntan_Zhang2005-2dnms} to allow variation with iteration. Simulation results on the (3096,1032) PBRL code show that N-2D-NMS decoder can achieve the same FER as N-NMS with significantly fewer parameters. 
  The N-2D-NMS decoder for a (16200,7200) DVBS-2 LDPC code achieves a lower error floor than belief propagation.
    \item  A hybrid decoding structure that combines a feedforwad structure with a recurrent structure that shows similar decoding performance as a full feedforward structure, but requires  significantly fewer parameters.

\end{itemize}

The remainder of the paper is organized as follows: Sec.~\ref{sec: effi_NNMS} derives the gradients of the loss function with respect to learnable parameters and neurons of a N-NMS neural network and proposes an efficient learning representation. Statistics of learned weights are studied in this section. Sec. \ref{sec: n2dnms} introduces a family of N-2D-NMS decoders. Sec. \ref{sec: Simulation} presents and discusses our  simulation results and explores a hybrid decoding structure. Sec. \ref{sec: conclusion} concludes our work.

\section{Efficient implementation of N-NMS}\label{sec: effi_NNMS}

\subsection{Forward Propagation}
Let $H$ be the parity check matrix of a $(n,k)$ LDPC code. We use $v_i$ and $c_j$ to denote the $i^{th}$ variable node and $j^{th}$ check node, respectively. N-NMS resembles NMS except assigning multiplicative parameters for each check-to-variable message in each iteration. In the $t^{th}$ \textit{decoding iteration}, N-NMS updates the check-to-variable node message, $u^{(t)}_{c_j \rightarrow v_i}$ , the variable-to-check node message, $l^{(t)}_{v_i\rightarrow c_j}$, and posterior of each variable node, $l_{v_i}^{(t)}$, by: 
\begin{align}
\begin{split}
       %u_{c_i\rightarrow v_j}&= \prod_{v_{j'}\in N(c_i)/\{v_j\}} \text{sgn}(l^{(t-1)}_{v_{j'}\rightarrow c_{i}}) \times \min_{v_{j'}\in N(c_i)/\{v_j\}} |(l^{(t-1)}_{v_{j'}\rightarrow c_{i}}|
    u^{(t)}_{c_i\rightarrow v_j} &= \beta^{(t)}_{(c_i,v_j)} \times  \prod_{v_{j'}\in N(c_i)/\{v_j\}} \text{sgn}(l^{(t-1)}_{v_{j'}\rightarrow c_{i}}) \\ & \times  \min_{v_{j'}\in N(c_i)/\{v_j\}} \left|(l^{(t-1)}_{v_{j'}\rightarrow c_{i}})\right| 
\end{split},
\\ 
l^{(t)}_{v_j\rightarrow c_i} &=  l^{ch}_{v_i} + \sum_{c_{i'}\in N(v_j)/\{c_i\}} u^{(t)}_{c_{i'}\rightarrow v_j},\\
l^{(t)}_{v_j} &= l^{ch}_{v_i} + \sum_{c_{i'}\in N(v_j)} u^{(t)}_{c_{i'}\rightarrow v_j}.
\end{align}
$N(c_i)$  ($N(v_j)$) represents the set of the variable nodes (check nodes) that are connected to $c_i$($v_j$). $l^{ch}_{v_j}$ is the LLR of channel observation of  $v_j$. $\beta^{(t)}_{(c_i,v_j)}$ are multiplicative weights to be trained. The decoding process stops when all parity checks are satisfied or maximum iteration $I_T$ is reached.

\subsection{Backward Propagation}

In this subsection, we derive the gradient of $J$ with respect to the learnable weights, $\frac{\partial J}{ \partial \beta^{(t)}_{(v_i,c_j)}}$, the check-to-variable message, $\frac{\partial J}{\partial u^{(t)}_{c_i\rightarrow v_j}}$, and variable-to-check message, $\frac{\partial J}{\partial l^{(t)}_{v_j \rightarrow u_i}}$. Just like forward propagation, $\frac{\partial J}{\partial u^{(t)}_{c_i\rightarrow v_j}}$ and  $\frac{\partial J}{\partial l^{(t)}_{v_j \rightarrow u_i}}$ are traced back iteratively. We show that in order to get desired gradients, it is sufficient only to store, $l_{v_i}^{(t)}$ ,sgn($l^{(t)}_{v_j \rightarrow c_i}$), sgn($u^{(t)}_{c_i\rightarrow v_j}$), $\texttt{min1}^{t}_{c_i}$, $\texttt{min2}^{t}_{c_i}$, $\texttt{pos1}^{t}_{c_i}$ and $\texttt{pos1}^{t}_{c_i}$ when performing forward propagation, where
\begin{align}
     \texttt{min1}^{t}_{c_i} &= \min_{v_{j'}\in N(c_i)} |l^{(t)}_{v_{j'}\rightarrow c_{i}}|\label{equ: min1}, \\
     \texttt{pos1}^{t}_{c_i} &= \argminB_{v_{j'}\in N(c_i)} |l^{(t)}_{v_{j'}\rightarrow c_{i}}|,\\
     \texttt{min2}^{t}_{c_i} &= \min_{v_{j'}\in N(c_i)/\{\texttt{pos1}^{t}_{c_i}\}}|l^{(t)}_{v_{j'}\rightarrow c_{i}}|,\\
     \texttt{pos2}^{t}_{c_i} &= \argminB_ {v_{j'}\in N(c_i)/\{\texttt{pos1}^{t}_{c_i}\}}|l^{(t)}_{v_{j'}\rightarrow c_{i}}|\label{equ: pos}.
\end{align}

In this paper, multi-loss cross entropy \cite{Nachmani2016-bs} is used as loss function. Denote loss by $J$ and assume all-zero codewords are transmitted,  it is straightforward to calculate $\frac{\partial J}{ \partial l^{(t)}_{v_i}}$:
\begin{align}
    \frac{\partial J}{ \partial l^{(t)}_{v_i}}= -\frac{1}{n I_T} \sigma(-l^{(t)}_{v_i}).
\end{align}
We initialize $\frac{\partial J}{\partial l^{(I_T)}_{vi\rightarrow c_j}}=0$ for $(c_i,v_j)$ pairs whose $H(i,j)=1$.

In iteration $t$,  $\frac{\partial J}{\partial l^{(t)}_{v_j \rightarrow u_i}}$ is updated based on  "back propagation through time" \cite{Werbos1988-cx}:
\begin{align}
\label{equ: update-c}
    \frac{\partial J}{\partial u^{(t)}_{v_j \rightarrow c_i}} = \frac{\partial J}{\partial l^{(t)}_{v_j}}+ \sum_{c_{i'}\in N(v_j)/\{c_i\}}\frac{\partial J}{\partial l^{(t)}_{c_{i'}\rightarrow v_j}}.
\end{align}
$\frac{\partial L}{\partial \beta^{(t)}_{c_i \rightarrow v_j}}$ is calculated by:
\begin{align}
    \frac{\partial J}{\partial \beta^{(t)}_{c_i \rightarrow v_j}} = u^{(t)*}_{c_i \rightarrow v_j}\frac{\partial J}{\partial u^{(t)}_{c_i\rightarrow v_j}},
\end{align}
where 
\begin{align}
    u^{(t)*}_{c_i\rightarrow v_j} &= \text{sgn}(u^{(t)*}_{c_i \rightarrow v_j})\times |{u^{(t)*}_{c_i \rightarrow v_j}}|, \\
    \text{sgn}(u^{(t)*}_{c_i \rightarrow v_j}) &=   \prod_{v_{j'}\in N(c_i)/\{v_j\}} \text{sgn}(l^{(t-1)}_{v_{j'}\rightarrow c_{i}}), \\
    |{u^{(t)*}_{c_i \rightarrow v_j}}| &= 
     \left\{ \begin{array}{l l}  \texttt{min2}^{t}_{c_i}, &  \quad  \text{if }v_j =  \texttt{pos1}^{t}_{c_i}   \\  \texttt{min1}^{t}_{c_i}, &  \quad \text{otherwise} \\ \end{array}. \right.
\end{align}

With chain rule, we can get $\frac{\partial J}{\partial |{u^{(t)*}_{c_i \rightarrow v_j}} |}$:
\begin{align}
    \frac{\partial J}{\partial |{u^{(t)*}_{c_i \rightarrow v_j}} |}&= \text{sgn}(u^{(t)*}_{c_i \rightarrow v_j}) \frac{\partial J} {\partial{u^{(t)*}_{c_i \rightarrow v_j}} }, \\
    &= \text{sgn}(u^{(t)*}_{c_i \rightarrow v_j}) \beta_{(c_i,v_j)}^{(t)}\frac{\partial J}{\partial {u^{(t)}_{c_i \rightarrow v_j}} }.
\end{align}

% Note that :
% \begin{align}
%     \frac{\partial }{\partial x_i}\min (x_1,...,x_n ) = 
%     \left\{ \begin{array}{l l}  1 &  \quad  \text{if } x_i \text{ is minimum}   \\ 0  &  \quad \text{otherwise}. \\ \end{array} \right. ,
% \end{align}
% and $\frac{\partial \text{sgn}(x)}{\partial  x} = 0$.
For all variable nodes connected to check node $c_i$, only $\texttt{pos1}^{(t)}_{c_i}$ and $\texttt{pos2}^{(t)}_{c_i}$ receive backward information, therefore, $\frac{\partial J}{\partial l^{(t-1)}_{v_j\rightarrow c_i}}$ can be computed as follows:
\begin{align}
\label{equ: l_update}
    \left\{ \begin{array}{l l} \text{sgn}(l^{(t-1)}_{v_j\rightarrow c_i})\sum_{v_{j'}\in N(c_i)/ \{v_j\}} \frac{\partial J}{\partial |{u^{(t)*}_{c_i \rightarrow v_{j'}}} |} & , \text{if } v_j = \texttt{pos1}^{(t)}_{c_i}   \\ \text{sgn}(l^{(t-1)}_{v_j\rightarrow c_i}) \frac{\partial J}{\partial \left|{u^{(t)*}_{ c_i \rightarrow  \texttt{pos1}^{(t)}_{c_i}}} \right|}  &  ,\text{if } v_j = \texttt{pos2}^{(t)}_{c_i} \\  0 & , \text{otherwise}  . \\ \end{array} \right. 
\end{align} 

Eq.\eqref{equ: update-c}-\eqref{equ: l_update} indicate that $\frac{\partial J}{\partial {u^{(t)}_{c_i \rightarrow v_j}} }$ and $\frac{\partial J}{\partial l^{(t)}_{v_j\rightarrow c_i}}$  are calculated in a message passing manner. As a result, the full N-NMS neural network is not necessarily to be built. Besid{\tiny {\tiny }}es, the neuron values in each hidden layer can be stored efficiently using Eq.\eqref{equ: min1}-\eqref{equ: pos}. Both solves memory issue faced by Tensorflow, as pointed in \cite{Lugosch_undated-wi}. The training work in this paper is done by C++ with only CPU computation nodes. Finally, we use stochastic gradient decent method with Adam optimizer to update $\beta^{(t)}_{(c_i,v_j)}$ in each \emph{training iteration}.
 
\subsection{Simulation Results}
In this subsection, we use the efficient implementation described above to train the weights of N-NMS for a (3096,1032) protograph-based raptor-like (PBRL) LDPC code. The code we use is taken from \cite{cls_tool} (in \cite{7045568}). Encoded bits $x$ are modulated by BPSK and transmitted through additive white Gaussian noise channel (AWGNC). The N-NMS decoder is flooding scheduled and maximum decoding iteration is 10. 

\begin{figure}[t] 
    \centering
  \subfloat[\label{fig: mean-t}]{%
       \includegraphics[width=0.50\linewidth]{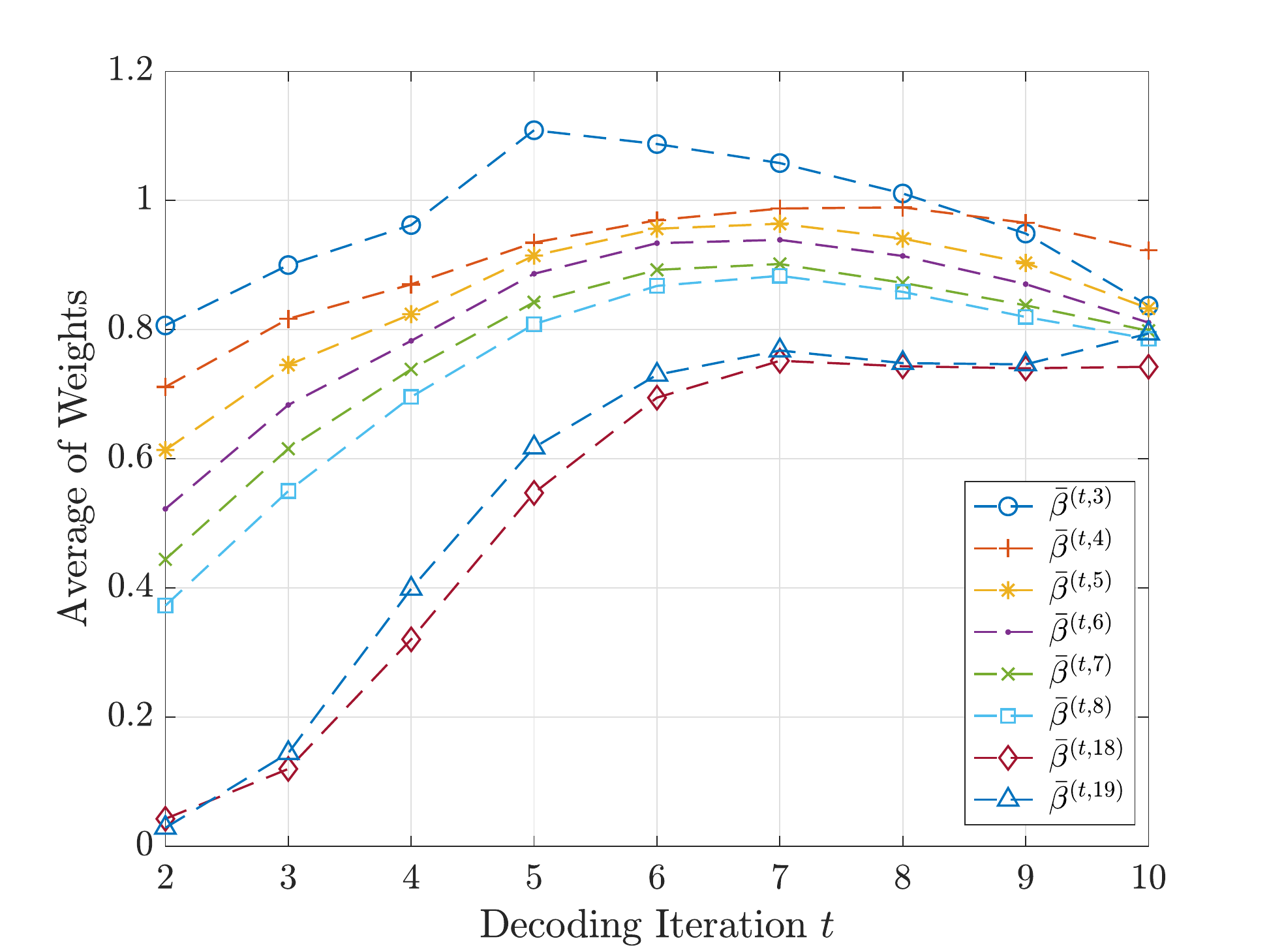}}
    \hfill
  \subfloat[\label{fig: degree-19-iter4}]{%
        \includegraphics[width=0.50\linewidth]{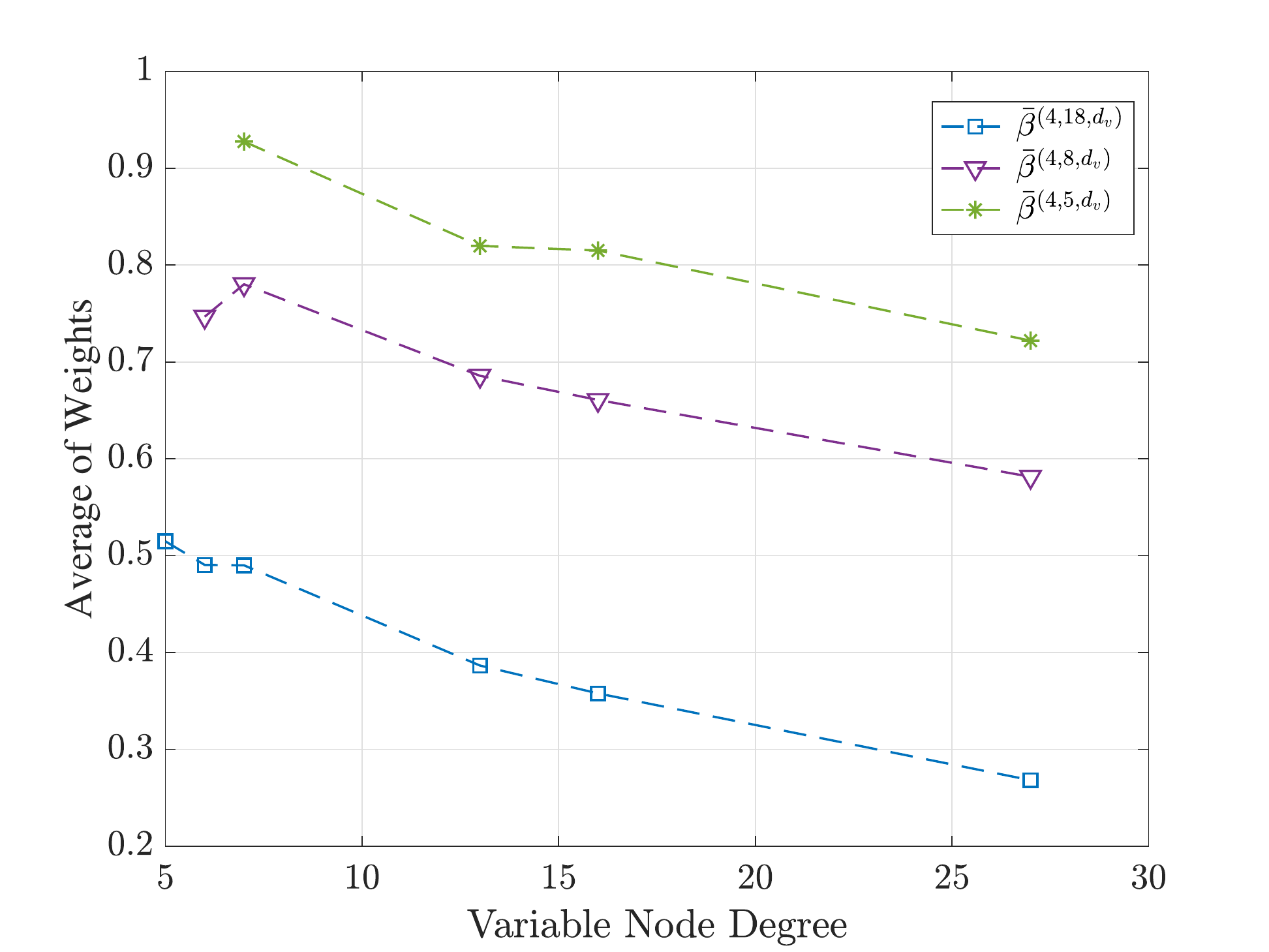}}
  \caption{Mean values of messages of FNNMS for a (3096,1032) PBRL code in each iteration show strong correlations to check and variable node degree.}
  \label{fig: gradient_explosion} 
\end{figure}

Define $\beta^{(t,d_c)}=\{\beta^{(t)}_{(c_i,v_j)}|\text{deg}(c_i) = d_c\}$ and $\bar{\beta}^{(t,d_c)}$ as the mean value of $\beta^{(t,d_c)}$. Fig.\ref{fig: mean-t} gives $\bar{\beta}^{(t,d_c)}$ versus decoding iteration $t$ with all possible check node degrees. Note that iteration number starts from 2 because most of edges have 0 messages in the first iteration, which is a result of puncturing.  The simulation shows a clear relationship between check node degree and $\bar{\beta}$, i.e. the larger check node degree corresponds to a  smaller $\bar{\beta}$, this difference is significant in the first few iterations.
Also, for each possible $d_c$, $\bar{\beta}^{(t,d_c)}$ changes significantly in first few iterations.  In short conclusion, Fig.\autoref{fig: mean-t} shows that $\beta^{(t)}_{(c_i,v_j)}$ has a strong correlation with check node degree and iteration.

In order to investigate the relationship between weights and variable node degree given a check node degree and iteration number, we further define $\beta^{(t,d_c,d_v)}=\{\beta^{(t)}_{(c_i,v_j)}| \text{deg}(c_i)=d_c,\text{deg}(v_i)=d_v\}$. We denote $\bar{\beta}^{(t,d_c,d_v)}$ by the average value of $\beta^{(t,d_c,d_v)}$. Fig.\ref{fig: degree-19-iter4} gives the average of weights corresponding to different check node degree and variable node degree in iteration 4. Simulation result shows that given a specific iteration $t'$ and check node degree $d_c'$, the larger $d_v'$ corresponds to  a smaller $\bar{\beta}^{(t',d_c',d_v')}$. 

In conclusion, the weights of N-NMS are correlated with check node degree, variable node degree and iteration. Thus, node degrees should affect the weighting of messages on their incident edges when decoding of irregular LDPC code. Inspired by recent neural network decoders, we propose a family of N-2D-NMS decoders in this paper. 

\section{Neural 2D Normalized MinSum Decoders }\label{sec: n2dnms}
Based on the previous discussion, it is intuitive to consider assigning same weights to messages with same check node degree and/or variable node degree. In this section, we propose neural 2-dimensional normalized MimSum (N-2D-NMS) decoders which has the following form:

\begin{align}
\begin{split}
       %u_{c_i\rightarrow v_j}&= \prod_{v_{j'}\in N(c_i)/\{v_j\}} \text{sgn}(l^{(t-1)}_{v_{j'}\rightarrow c_{i}}) \times \min_{v_{j'}\in N(c_i)/\{v_j\}} |(l^{(t-1)}_{v_{j'}\rightarrow c_{i}}|
    u^{(t)}_{c_i\rightarrow v_j} &= \beta^{(t)}_{*} \times  \prod_{v_{j'}\in N(c_i)/\{v_j\}} \text{sgn}(l^{(t-1)}_{v_{j'}\rightarrow c_{i}}) \\ & \times  \min_{v_{j'}\in N(c_i)/\{v_j\}} |(l^{(t-1)}_{v_{j'}\rightarrow c_{i}})| 
\end{split},
\\ 
l^{(t)}_{v_j\rightarrow c_i} &=  l^{ch}_{v_i} + \alpha^{(t)}_{*}  \sum_{c_{i'}\in N(v_j)/\{c_i\}} u^{(t)}_{c_{i'}\rightarrow v_j}.
\end{align}

The $\beta^{(t)}_{*}$ and $\alpha^{(t)}_{*}$ are the multiplicative weights assigned to check and variable node messages, respectively. The subscripts indicate weight sharing metric. Table. \ref{tab:weight_sharing} lists different types of N-2D-NMS decoders. As a special case, we denote NNMS as type-0.

Type-1 to type-4 assign the same weights based on node degree. Type-1 N-2D-NMS assigns same weights to the edges that have same check node \emph{and} variable node degree. Type-2 considers the check node degree and variable node degree separately. As a simplification, type-3 and type-4 only consider variable node degree and check node degree, respectively.

Dai. \emph{et. al} in \cite{Dai2021-ou} studied weight sharing based on edge type of MET-LDPC code, or protograph-based code. We also consider this metric in paper, which is given by type-5, 6 and 7. Type-5  assigns same weights to the edges with same edge-type, i.e. the edges belonged to the same position in protomatrix. $f$ is lifting factor. In this paper, we further consider its simplifications, type-6 and-type 7, which assign parameters only based on horizontal layers and vertical layers, respectively.
Finally, type 8 decoder assigns iteration-distinct parameters, this simple weight sharing schemes have been considered for previous literature \cite{Lian2019-jh,Abotabl2019-wt}. 

Table.\ref{tab:weight_sharing} gives the number of parameters per iteration required for various N-2D-NMS decoders. The (3096,1032) PBRL code and (16200,7200) DVBS-2\cite{noauthor_2019-nv} standard LDPC code are considered in this paper. It is shown that  the number of parameters required by node-degree based weight sharing is less than that required by protomatrix based weight sharing. Based on the simulation results given in Sec. \ref{sec: Simulation}, the two weight sharing schemes deliver same error correction performance.

\begin{center}
    \begin{table}[t]
\makegapedcells
\caption{\label{tab:weight_sharing} Various N-2D-NSM Decoders and \\ Required Number of Parameters Per Iteration}
\begin{tabular}{|c|c|c|c|c|}
\hline
type & $\beta^{(t)}_{*}$ & $\alpha^{(t)}_{*}$ &\begin{tabular}[c]{@{}c@{}}(16200,7200) \\ DVBS-2 code\end{tabular} & \begin{tabular}[c]{@{}c@{}}(3096,1032) \\ PBRL code\end{tabular}    \\ \hline
$0^{[1]}$ & $\beta^{(t)}_{(c_i,v_j)}$ & 1 & $4.8*10^5$ & $1.60*10^4$ \\ \hline
\multicolumn{5}{|c|}{Weight Sharing Based on Node Degree} \\ \hline
1 &  $\beta^{(t)}_{(d.{(c_i)},d.{(v_j)})}$   & 1 & 13 & 41 \\ \hline
2 & $\beta^{(t)}_{(deg{(c_i)})}$ & $\alpha^{(t)}_{(deg{(v_j)})}$ & 8 & 15 \\ \hline
3 & $\beta^{(t)}_{(deg{(c_i)})}$ & 1 & 4 & 8 \\ \hline
4 & 1 & $\alpha^{(t)}_{(deg{(v_j)})}$ & 4 & 7 \\ \hline
\multicolumn{5}{|c|}{Weight Sharing Based on Protomatrix} \\ \hline
$5^{[20]}$ & $\beta^{(t)}_{\left( \floor{\frac{i}{f}},\floor{\frac{j}{f}}\right)}$ & 1 & $-$ & 101 \\ \hline
6 & $\beta^{(t)}_{\left( \floor{\frac{i}{f}}\right)}$ & 1 & $-$ & 17 \\ \hline
7 & 1 & $\beta^{(t)}_{\left( \floor{\frac{j}{f}}\right)}$ & $-$ & 25 \\ \hline
\multicolumn{5}{|c|}{Weight sharing based on Iteration \cite{Lian2019-jh,Abotabl2019-wt}} \\ \hline
8 & $\beta^{(t)}$ & 1 & 1 & 1 \\ \hline
\end{tabular}
\end{table}
\end{center}

\section{Simulation Result}\label{sec: Simulation}

In this section, we investigate the decoding performance of N-2D-NMS decoders for the LDPC codes with different block length. All encoded bits are modulated by BPSK and transmitted through AWGNC. The LDPC codes and optimized weights in this paper can be found and downloaded in \cite{cls_tool}.
 \begin{figure}[t]
	\centering
	 \includegraphics[width=20pc]{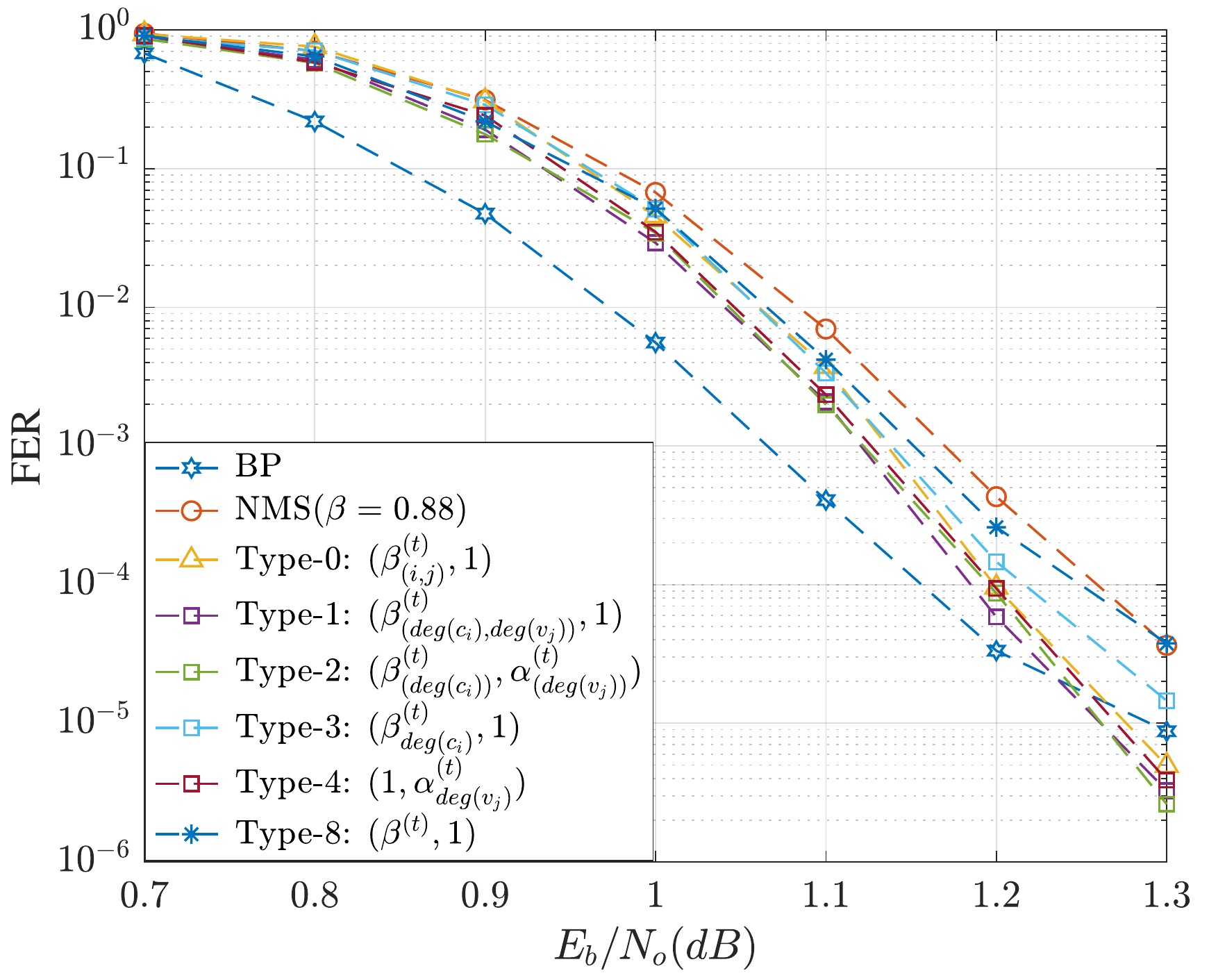}
	\caption{FER performance N-2D-NMS decoders for a (3096,1032) PBRL LDPC code. }
    \label{fig: dvbs2-fer} 
\end{figure}
\begin{figure}[t] 
\centering
  \subfloat[{$\beta^{(t)}_{deg(c_i)}$ and $\alpha^{(t)}_{deg(v_j)}$ of type-2 N-2DNMS decoder.}\label{fig: meanvalue-type2}]{%
       \includegraphics[width=1.00\linewidth]{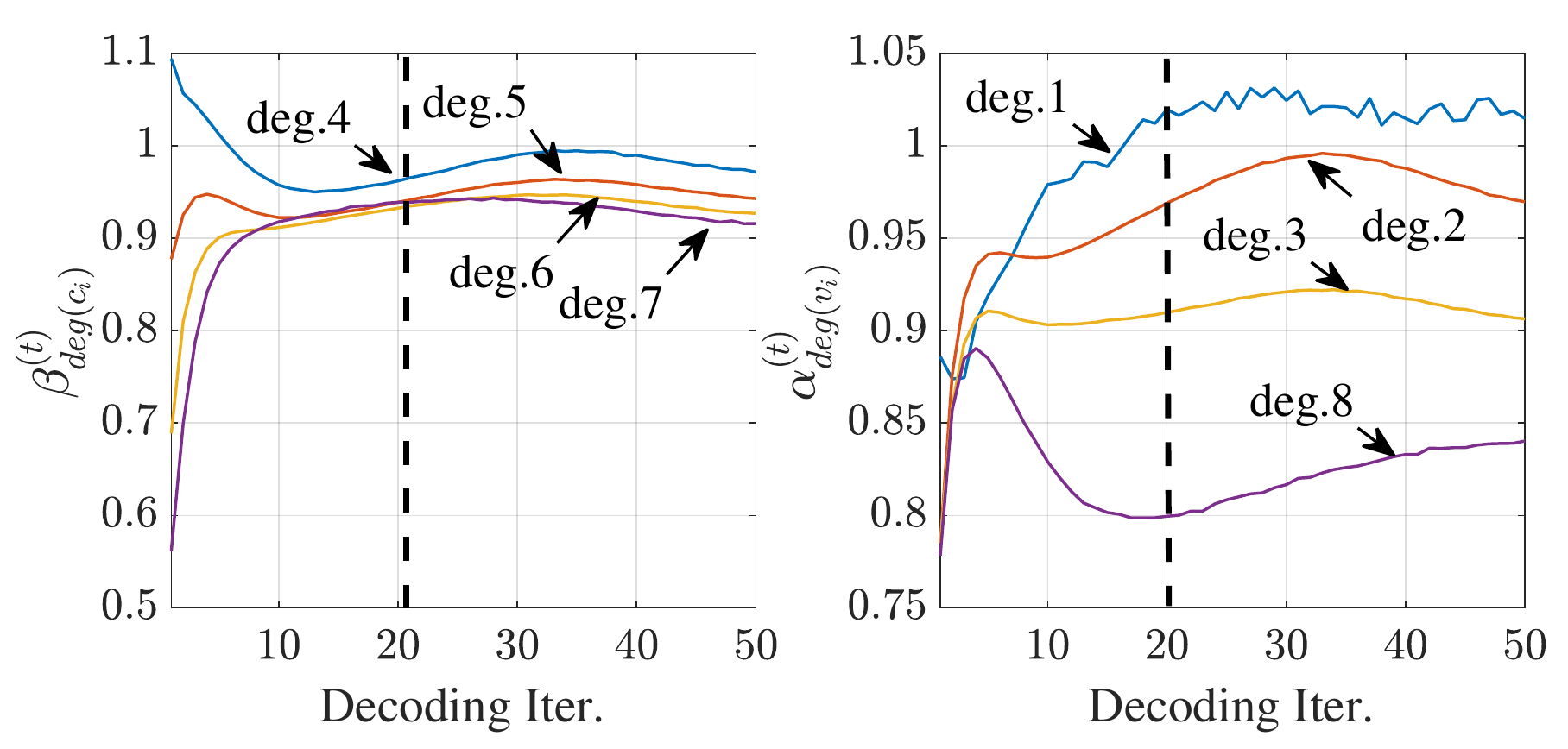}}
        
  \subfloat[$\beta^{(t)}$ of type-8 N-2D-NMS. \label{fig: meanvalue-type8}]{%
        \includegraphics[width=0.50\linewidth]{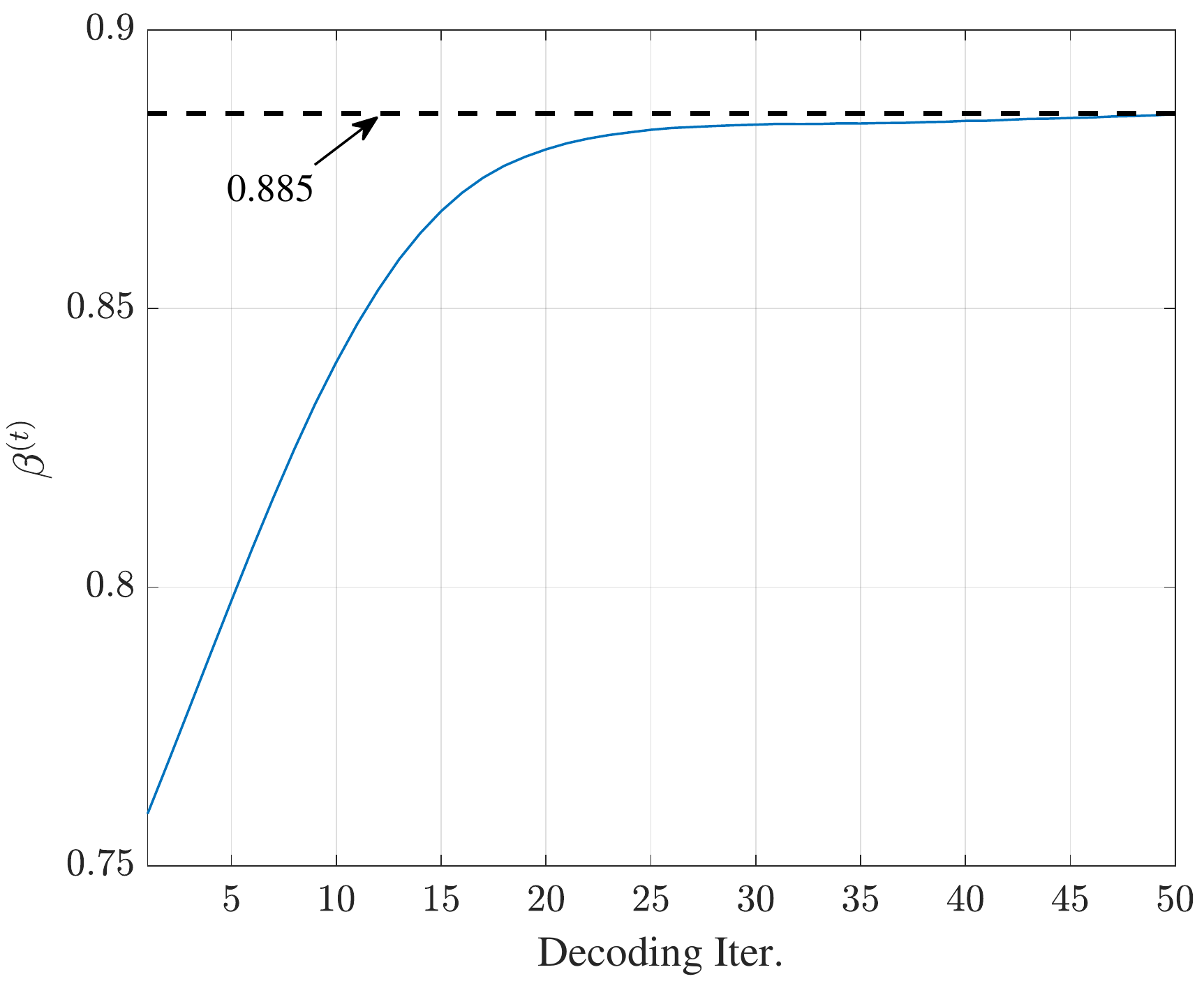}}
   \hfill
  \subfloat[FER performance of hybrid type-2 N-2D-NMS decoder\label{fig: dvbs2-fer2}]{%
        \includegraphics[width=0.50\linewidth]{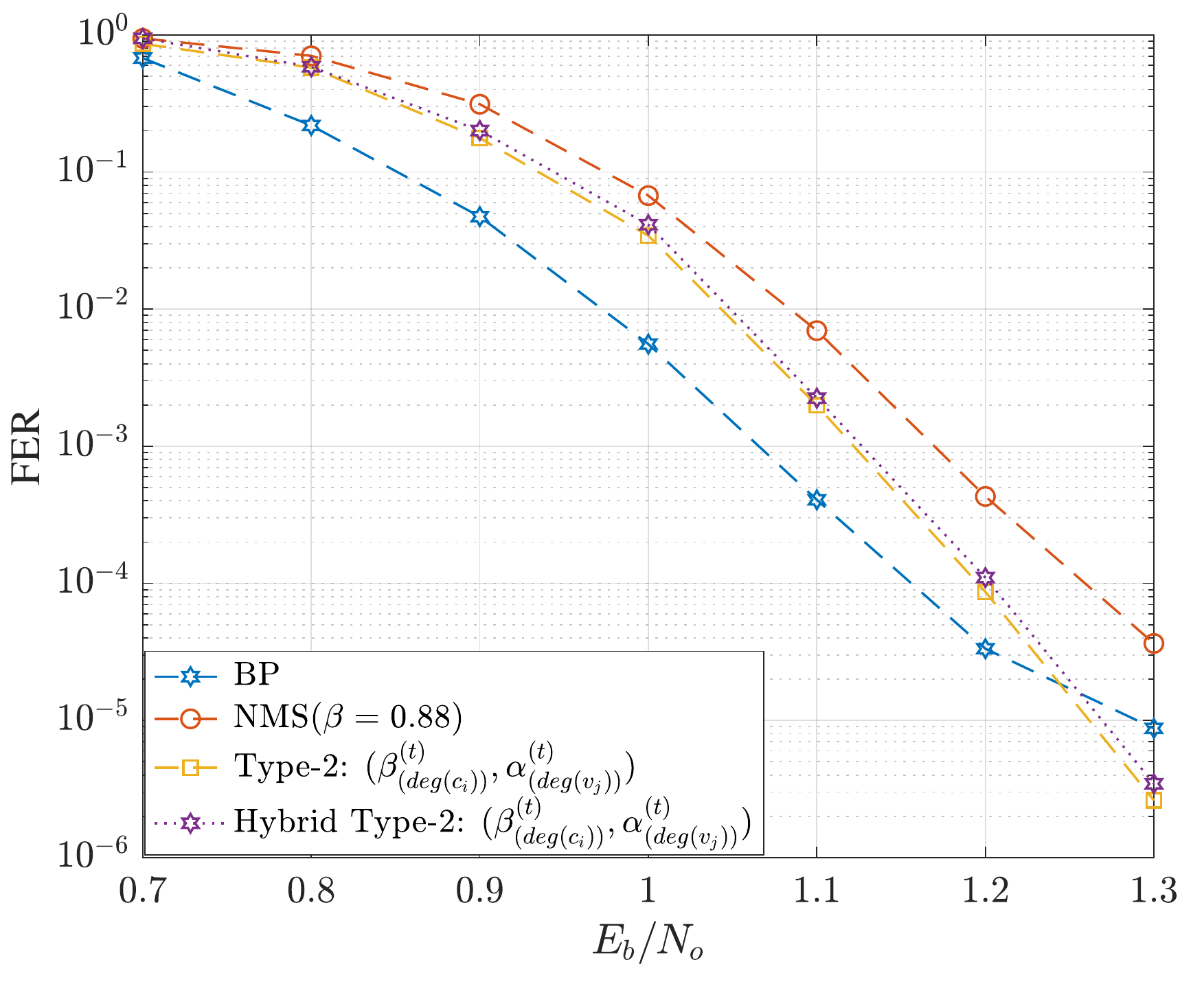}}
 \caption{Trained weights of N-2D-NMS decoders for a (16200,7200) DVBS-2 code only changes in significantly in the first 20 iterations. The hybrid type-2 N-2D-NMS decoder with $I'=20$ shows a comparable decoding performance to type-2 decoder, with 60\% parameters reduction. }
  \label{fig: weight_change}
  \hfill
\end{figure}
\subsection{(16200,7200) DVBS-2 LDPC code}

Fig. \ref{fig: dvbs2-fer} gives the FER performances of various LDPC decoder for (16200,7200) DVBS-2 LDPC code. 
All the decoders are implemented in a \emph{flooding way} and maximum decoding iteration is 50. 
It is shown that N-NMS decoder outperforms BP at $1.3dB$, with a lower error floor.
Type-1 and type-2 N-2D-NMS decoders have a slightly better performance than N-NMS. 
As two simplifications of type-2 N-2D-NMS, type 4 decoder outperforms type 3, because the variable node weights of investigated code have a larger dynamic range than check node weights, as shown in Fig.\ref{fig: meanvalue-type2}.

Fig. \ref{fig: meanvalue-type2} shows the $\beta^{(t)}_{(\text{deg}({c_i}))}$ and $\alpha^{(t)}_{(\text{deg}({v_j}))}$ of type-2 decoder. The trained values agree with our observation in the previous section, i.e., in each decoding iteration, larger degree node corresponds to a smaller value.  Note that the parameters only changes greatly in the first 20 iterations, and after that the change is minor.  It directly leads to a combination of a feedforward neural network with a recurrent neural network, meaning that parameters are updated in the first $I'$ iterations and the last $I_T-I'$ iterations reuse the parameters in iteration $I'$.  We call this structure as \emph{hybrid} N-2D-NMS decoder. Fig. \ref{fig: dvbs2-fer2} shows that hybrid type-2 N-2D-NMS wiht $I'_T=20$ has comparable decoding performance with full feedforward decoding structure, with 60\% parameter reduction.

Fig. \ref{fig: meanvalue-type8} shows that the parameters of type-8 decoder converge to 0.885, which is close to the single weight of  NMS decoder. As shown in Fig.\ref{fig: dvbs2-fer}, by only assigning iteration-specific parameters, type-8 decoder appears an early error floor at $1.20dB$.

\subsection{(3096,1032) PBRL LDPC Code}

\begin{figure}[t]
	\centering
	\includegraphics[width=20pc]{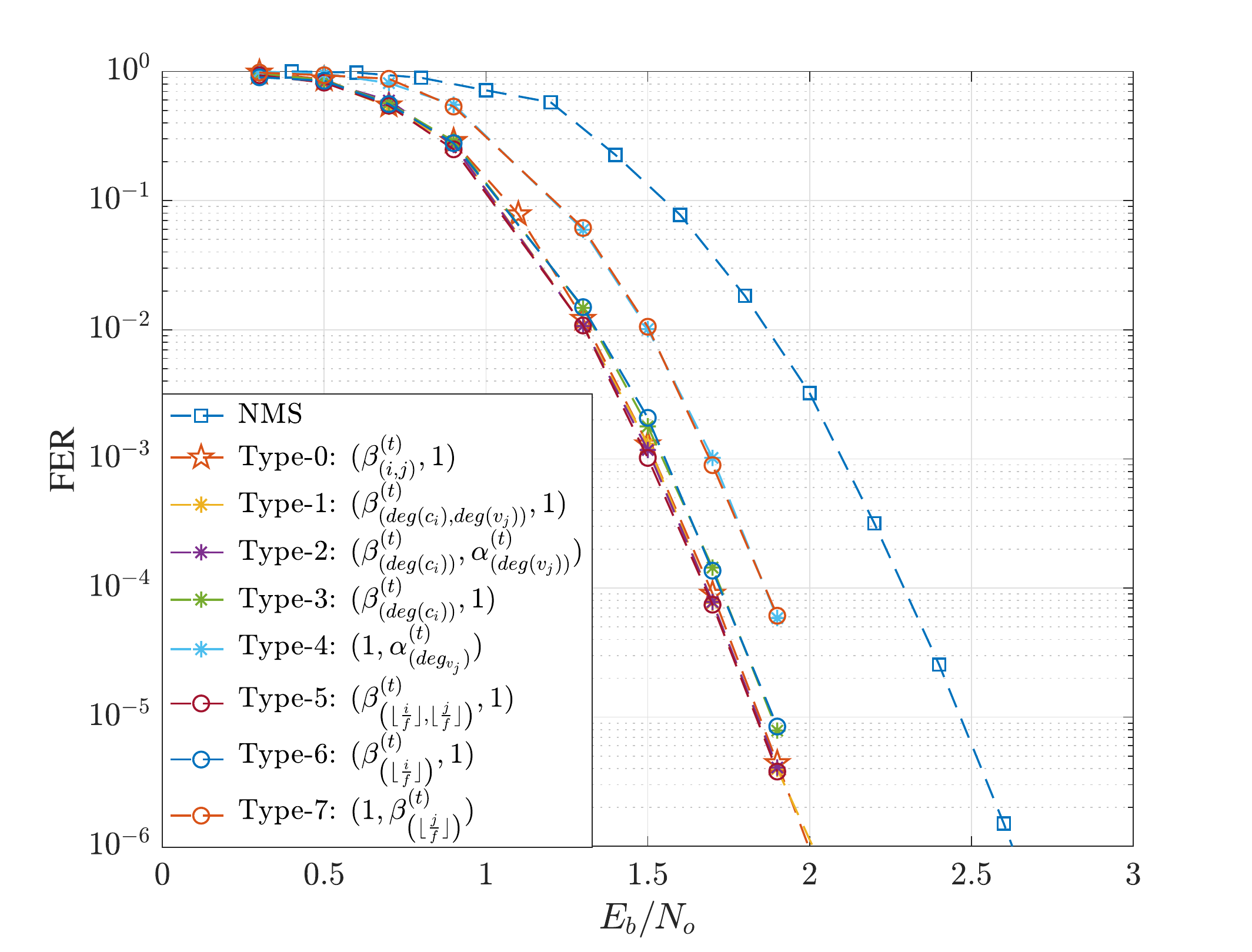}
	\caption{FER performance N-2D-NMS decoders for a (3096,1032) PBRL LDPC code. }
    \label{fig: FER1}
\end{figure}

Fig.\ref{fig: FER1}  gives the FER performance of N-2D-NMS decoders. All the decoder are implemented in a \emph{layered way} and maximum decoding iteration is 10. The simulation results show that NNMS (i.e. type 0) has a more than 0.5$dB$ improvement than NMS decoder. Type 1-7 decoders are also simulated. Note that type 1,2 and 5 have same performance with NNMS decoder, with much smaller number of parameters required. It is shown that weight sharing metrics based on check and variable node degree, or based on horizontal and vertical layer delivers lossless performance.  Type-4 and 6 decoders have a degradation around 0.05$dB$ compared with type 0, type-5 and 7 have a degradation around 0.2$dB$ compared type 0. It can be seen that, for (3096,1032) PBRL code, assigning weights only based on check nodes can gain more benefit than assigning weights only based on variable node.

\section{Conclusion}\label{sec: conclusion}

This paper presents MinSum LDPC decoders for which the normalization weights are optimized by a neural network.  An initial neural network allows a different weight for every edge.  The statistics of the trained parameters show a strong correlation with check node degree, variable node degree and iteration. A family of neural 2D normized MinSum (N-2D-NMS) decoders are introduced in this paper with different parameter reductions and decoding performances.
Simulation results on a (16200,7200) DVBS-2 standard LDPC code and (3096,1032) PBRL code show that N-2D-NMS decoder can have same decoding performance as NNMS decoder and achieves a lower error floor than BP.
Finally, a hybrid decoding structure combining feedforward structure with recurrent structure is proposed in this paper. The hybrid structure shows similar decoding performance to full feedforward structure, with less parameters required.

\bibliographystyle{IEEEtran}
\bibliography{nn_decoder,csl}

\end{document}